\documentclass[preprint,aps,prl]{revtex4}
\usepackage[dvips]{graphicx}
\usepackage{color,array,dcolumn}
\usepackage{amsmath}
\usepackage{amssymb}

\begin{document}

\title{van der Waals-corrected Density Functional Theory
simulation of adsorption processes on noble-metal surfaces:
Xe on Ag(111), Au(111), and Cu(111)}

\author{Pier Luigi Silvestrelli and Alberto Ambrosetti} 
\affiliation{Dipartimento di Fisica e Astronomia, 
Universit\`a di Padova, via Marzolo 8, I--35131, Padova, Italy,
and DEMOCRITOS National Simulation Center, of the Italian Istituto 
Officina dei Materiali (IOM) of the Italian National 
Research Council (CNR), Trieste, Italy}

\begin{abstract}
\date{\today}
The DFT/vdW-WF2s1 method based on the generation of localized Wannier 
functions, recently developed to include the van der Waals
interactions in the Density Functional Theory and describe adsorption
processes on metal surfaces by taking metal-screening effects into
account, is applied to the case of the interaction of Xe 
with noble-metal surfaces, namely Ag(111), Au(111), and Cu(111).
The study is also repeated by adopting the DFT/vdW-QHO-WF
variant relying on the Quantum Harmonic Oscillator model which 
describes well many-body effects. 
Comparison of the computed equilibrium binding energies and distances,
and the $C_3$ coefficients characterizing the
adatom-surface van der Waals interactions, 
with available experimental and theoretical
reference data shows that the methods perform well and elucidate the
importance of properly including screening effects. 
The results are also compared with those obtained by 
other vdW-corrected DFT schemes, including PBE-D, vdW-DF, vdW-DF2, rVV10,
and by the simpler Local Density Approximation (LDA) and semilocal (PBE)
Generalized Gradient Approximation (GGA) approaches.
\end{abstract}

\maketitle

\section{Introduction}
Adsorption processes on solid surfaces represent a very important topic
both from a fundamental point of view and to design and optimize 
countless material applications.
In particular, the adsorption of rare-gas atoms on metal 
surfaces is prototypical\cite{Bruch} for ''physisorption'' processes,
characterized by an equilibrium between attractive, 
long-range van der Waals (vdW) interactions and short-range Pauli repulsion
acting between the electronic charge densities of the substrate 
and the adsorbed atoms and molecules\cite{Vidali}.

Rare-gas adsorption on many close-packed metal surfaces, such
as Ag(111), Al(111), Cu(111), Pd(111), Pt(111),.. have been 
extensively studied both
experimentally\cite{Gottlieb,Seyller,Narloch,Diehl} and 
theoretically\cite{Diehl,Silva,DaSilva05,DaSilva,Betancourt,Lazic,Righi,Sun,Chen,Chen12}.
Due to the non-directional character of the vdW
interactions that should be dominant in physisorption processes, 
surface sites that maximize the coordination of the rare-gas adsorbate atom 
are expected to be the preferred ones, thus favoring the {\it hollow}
adsorption site. However, 
recent studies indicate that the actual scenario is 
more complex: in particular, for Xe a general tendency is 
found\cite{Diehl,DaSilva05,DaSilva,Betancourt,Lazic} for
adsorption on metallic surfaces in the low-coordination {\it top} sites
(this behavior is attributed\cite{Diehl,Bagus} to the delocalization
of charge density that increases the Pauli repulsion effect at the
{\it hollow} sites relative to the {\it top} site and lifts the potential well
upwards both in energy and height).

Density Functional Theory (DFT) is a well-established
computational approach to study
the structural and electronic properties of
condensed matter systems from first principles, and, in particular, to
elucidate complex surface processes such
as adsorptions, catalytic reactions, and diffusive motions. 
Although current density functionals are able to describe quantitatively
condensed matter systems at much lower computational cost than other
first principles methods, they fail\cite{Kohn} to properly describe
dispersion interactions. Dispersion forces originate from 
correlated charge oscillations in
separate fragments of matter and the most important 
component is represented by the $R^{-6}$ vdW interaction\cite{london},
originating from correlated instantaneous dipole
fluctuations, which plays a fundamental role
in adsorption processes of fragments weakly interacting with a substrate
(''physisorbed'').

This is clearly the case for the present systems which can be divided 
into well separated fragments (adatoms and the metal substrate)
with negligible electron-density overlap.
The local or semilocal character 
of the most commonly employed exchange-correlation functionals 
makes DFT methods unable to correctly predict binding energies and 
equilibrium distances within both the local density (LDA) and the 
Generalized Gradient approximations (GGA)\cite{Riley}.
Typically, in many physisorbed systems GGAs give only a shallow and 
flat adsorption well at large adparticle-substrate separations, while 
the LDA binding energy often turns out to be not far from the experimental 
adsorption energy; however, since it is well known that LDA 
tends to overestimate the binding in systems with 
inhomogeneous electron density (and to underestimate the equilibrium
distances), the reasonable 
performances of LDA must be considered as accidental.
Therefore, a theoretical approach beyond the
DFT-LDA/GGA framework, that is able to properly describe vdW effects 
is required to provide more quantitative results\cite{DaSilva}.

In the last few years a variety of practical methods have been proposed
to make DFT calculations able to accurately describe vdW effects (for a
recent review, see, for instance, refs. \cite{Riley,MRS,Klimes}).
We have developed a family of such methods, all based on the generation
of the Maximally Localized Wannier Functions (MLWFs)\cite{Marzari},
successfully applied to a variety of
systems, including small molecules, water clusters,
graphite and graphene, water layers interacting with graphite,
interfacial water on semiconducting substrates,
hydrogenated carbon nanotubes,
molecular solids, and the interaction of rare gases and small molecules
with metal surfaces\cite{silvprl,silvmetodo,silvsurf,CPL,silvinter,ambrosetti,Costanzo,Ar-Pb,Ambrosetti2012,C3,PRB2013,QHO-WF,QHO-WFs}.
Of a particular value is the possibility of dealing with metal surfaces;
in fact insulating surfaces could be somehow treated even using atom-based
semiempirical approaches where an approximately derived $R^{-6}$ term,
multiplied by a suitable short-range damping function, is 
explicitly introduced. 
Instead, in our methods the atom-based point of view assumed in standard
semiempirical approaches is replaced by an electron-based point of view,
so that the schemes are also applicable to systems, such as metals and
semimetals, which cannot be described in terms of assemblies of atoms only
weakly perturbed with respect to their isolated configuration.
                                              
In particular, the DFT/vdW-WF2s1 method, presented in
ref. \cite{PRB2013}, has been specifically
developed to take metal-screening effects into account
and has been applied to the study of the
adsorption of rare gases and small molecules on different metal surfaces,
namely Al(111), Cu(111), and Pb(111), which are systems where a proper
inclusion of screening is essential\cite{Ruiz}.
In fact the vanishing band gap of the metal substrate leads to a 
fully non-local collective substrate response that effectively screens the 
interactions, thereby significantly reducing effective $C_6$ coefficients and 
polarizability.

The DFT/vdW-QHO-WF variant\cite{QHO-WF,QHO-WFs} combines
the Quantum Harmonic Oscillator (QHO) model with the MLWFs, in such a
way to be no longer restricted to the case of well separated
interacting fragments and to include higher than pairwise energy contributions,
coming from the dipole--dipole coupling among quantum oscillators.
DFT/vdW-QHO-WF hence provides a more complete description of the long-range
correlation energy, beyond second-order London dispersion. In particular,
the QHO model naturally accounts for non-additive long-range many body
effects\cite{QHO-WF,QHO-WFs}, deriving from the
self-consistent screening of the system polarizability induced by the
Coulomb interaction.
In the specific case of adsorption on metal surfaces a long-range damping
factor has been introduced\cite{QHO-WFs} to take metal-screening effects
into account.

We have already investigated, by the first version of the approach
based on the use of the MLWFs,
that is the DFT/vdW-WF method\cite{silvprl,silvmetodo,Mostofi}, 
the adsorption of Xe on the Cu(111) and Pb(111) surfaces\cite{Ar-Pb},
however in those applications a more crude description of screening effects
in metal substrates was adopted. 

We here apply our more recent schemes mentioned above, namely
DFT/vdW-WF2s1 and DFT/vdW-QHO-WF, 
to the case of adsorption of Xe on the Ag(111), Au(111), 
and Cu(111) metal surfaces.
Our results will be compared to the best available, reference experimental
and theoretical values, and to those obtained by other DFT
vdW-corrected schemes, including dispersion corrected PBE (PBE-D\cite{PBE-D}),
vdW-DF\cite{Dion,Langreth07}, vdW-DF2\cite{Lee-bis}, rVV10\cite{Sabatini}
(the revised, more efficient version of the original VV10 scheme\cite{Vydrov}),
and by the simpler Local Density Approximation (LDA) and semilocal GGA
(in the PBE flavor\cite{PBE}) approaches.
In the PBE-D scheme DFT calculations at the PBE level are corrected
by adding empirical $C_6/R^6$ potentials with parameters derived from
accurate quantum chemistry calculations for atoms, while in other methods,
such as vdW-DF, vdW-DF2, and rVV10, vdW effects are included by
introducing DFT nonlocal correlation functionals.

\section{Method}
Basically (additional details can be found in refs. \cite{C3,PRB2013}), 
the DFT/vdW-WF2s1 method relies on the    
well known London's expression\cite{london} where
two interacting atoms, $A$ and $B$, are approximated by 
coupled harmonic oscillators
and the vdW energy is taken to be the change of the zero-point energy
of the coupled oscillations as the atoms approach; if only a single
excitation frequency is associated to each atom, $\omega_A$, $\omega_B$,
then

\begin{equation}
E^{London}_{vdW}=-\frac{3e^4}{2m^2}\frac{Z_A Z_B}{\omega_A \omega_B(\omega_A+\omega_B)}\frac{1}{R_{AB}^6}
\label{lond}
\end{equation}

where $Z_{A,B}$ is the total charge of $A$ and $B$, and $R_{AB}$ is 
the distance between the two atoms ($e$ and $m$ are the electronic charge
and mass).

Now, adopting a simple classical theory of the atomic polarizability, 
the polarizability 
of an electronic shell of charge $eZ_i$ and mass $mZ_i$, tied to a heavy 
undeformable ion can be written as

\begin{equation}
\alpha_i\simeq \frac{Z_i e^2}{m\omega_i^2}\,.
\label{alfa}
\end{equation}

Then, given the direct relation between polarizability and atomic
volume\cite{polvol}, we assume that $\alpha_i = \gamma S_i^3$,
where $\gamma$ is a proportionality constant, so that the atomic volume is
expressed in terms of the MLWF spread, $S_i$.
Rewriting Eq. \eqref{lond} in terms of the quantities defined above,
one obtains an explicit expression for the $C_6$ vdW coefficient:

\begin{equation}
C_{6}^{AB}=\frac{3}{2}\frac{\sqrt{Z_A Z_B}S_A^3 S_B^3 \gamma^{3/2}}
{(\sqrt{Z_B}S_A^{3/2}+\sqrt{Z_A}S_B^{3/2})}\,.
\label{c6}
\end{equation}

The constant $\gamma$ can then be set up by imposing that the exact value for
the H atom polarizability
($\alpha_H=$4.5 a.u.) is obtained; of course, in the H case, one
knows the exact analytical spread, $S_i=S_H=\sqrt{3}$ a.u.

In order to achieve a better accuracy, one must properly deal with
{\it intrafragment} MLWF charge overlap. 
This overlap affects the effective orbital volume,
the polarizability, and the excitation
frequency (see Eq. \eqref{alfa}), thus leading to a quantitative effect on the
value of the $C_6$ coefficient.
We take into account the effective change in volume due to intrafragment
MLWF overlap by introducing a suitable reduction factor $\xi$
obtained by interpolating between the limiting cases of fully
overlapping and non-overlapping MLWFs (see ref. \cite{C3}).
We therefore arrive at the following expression for the $C_6$ coefficient:

\begin{equation}
C_{6}^{AB}=\frac{3}{2}\frac{\sqrt{Z_A Z_B}\xi_A S_A^3 \xi_B S_B^3 \gamma^{3/2}}
{(\sqrt{Z_B\xi_A} S_A^{3/2}+\sqrt{Z_A\xi_B} S_B^{3/2})}\,,
\label{c6eff}
\end{equation}

where $\xi_{A,B}$ represents the ratio between the effective and the
free volume associated to the $A$-th and $B$-th MLWF.

Finally, the vdW interaction energy is computed as:

\begin{equation}
E_{vdW}=-\sum_{i<j}f(R_{ij})\frac{C_6^{ij}}{R^6_{ij}} \,,
\label{EvdW}
\end{equation}

where $f(R_{ij})$ is a short-range damping function defined as follows:

\begin{equation}
f(R_{ij})=\frac{1}{1+e^{-a(R_{ij}/R_s-1)}}\,.
\end{equation}

This short-range damping function is
introduced not only to avoid the unphysical divergence of the
vdW correction at small fragment separations, but also
to eliminate double countings of correlation effects, by considering that
standard DFT approaches properly describe short-range
correlations.

The parameter $R_s$ represents
the sum of the vdW radii $R_s=R_i^{vdW}+R_j^{vdW}$, with
(by adopting the same criterion chosen above for
the $\gamma$ parameter)
\begin{equation}
R_i^{vdW}=R_H^{vdW}\frac{S_i}{\sqrt{3}}
\end{equation}
where $R_H^{vdW}$ is the literature\cite{Bondi} (1.20 \AA) vdW radius of
the H atom, and, following Grimme {\it et al.}\cite{Grimme},
$a \simeq 20$; 
note that the results are only mildly
dependent on the particular value of this parameter, at least within a
reasonable range around the $a=20$ value: for instance,
the binding energy at the optimal Xe-Ag(111) distance changes by less
than 5\% by varying $a$ between 10 and 30.
Although the damping function introduces a certain degree of empiricism
in the method, we stress that $a$
is the only ad-hoc parameter present in our approach, while all the others
are only determined by the basic information given by the MLWFs, namely
from first principles calculations so that they adapt to the specific chemical
environment.

To get an appropriate inclusion of metal screening effects
a proper reduction coefficient is included by multiplying the
$C_6^{ij}/R^6_{ij}$ contribution in Eq. \eqref{EvdW} by a Thomas-Fermi factor:
$ f_{TF} = e^{-2(z_s-z_l)/r_{_{TF}}}$
where $r_{_{TF}}$ is the Thomas-Fermi screening length relative
to the electronic density of an effective uniform electron 
gas (''jellium model'') describing the substrate,
$z_s$ is the average vertical position of the topmost metal
atoms, and $z_l$ is the vertical coordinate of the Wannier Function Center
(WFC) belonging to the
substrate ($l=i$ if it is the $i$-th WFC which belongs to the
substrate, otherwise $l=j$); the above $f_{TF}$ function is only applied if
$z_l < z_s$, otherwise it is assumed that $f_{TF}=1$ (no screening effect). 

An alternative, even simpler approach to mimic screening effects in 
adsorption processes is represented by the so-called ''single-layer'' 
approximation, in which vdW effects are only restricted to the interactions 
of the adparticle with the topmost metal layer\cite{Hanke}; in fact,
as a consequence of screening, one expects that the
topmost metal atoms give the dominant contribution.
We have implemented this by multiplying the
$C_6^{ij}/R^6_{ij}$ factor in Eq. \eqref{EvdW} by a damping
function:
        
\begin{equation}
f_{SL} = 1-{\frac {1}{1 + e^{(z_l-z_r)/{\Delta z}}} } \;,
\end{equation}
       
where $z_l$ is the vertical coordinate of the WFC belonging to
substrate (again $l=i$ if it is the $i$-th WFCs which belongs to the
substrate, otherwise $l=j$), the reference level $z_r$ is
taken as intermediate between the level of the first, topmost
surface layer and the second one, and we assume that
$\Delta z = $(interlayer separation)$/4$; we found that the estimated
equilibrium binding energies and adparticle-surface distances exhibit
only a mild dependence on the $\Delta z$ parameter.
Clearly this approach, denoted as DFT/vdW-WF2s3\cite{PRB2013},
resembles the DFT/vdW-WF2s1 scheme, the
basic difference being that the Thomas-Fermi damping function of
DFT/vdW-WF2s1 is here replaced by the $f_{SL}$ damping function
introduced to just select the WFCs around the topmost
surface layer.

In the DFT/vdW-QHO-WF variant (further details can be
found in refs. \cite{QHO-WF,QHO-WFs}) one exploits instead 
the fact that
for a system of $N$ three-dimensional QHOs the exact total energy can be 
obtained\cite{Cao,Donchev,Tkatchenko12,Reilly,QHO} 
by diagonalizing the $3N \times 3N$ 
matrix $C^{QHO}$, containing $N^2$ blocks $C_{ij}^{QHO}$ of size $3 \times 3$:

\begin{equation}
C_{ii}^{QHO} = \omega_i^2{\bf I}\,\,\,\, ; \,\,\,\,
C_{i\neq j}^{QHO} = \omega_i\omega_j{\sqrt {\alpha_i\alpha_j}}T_{ij}
\label{CQHO}
\end{equation}

where ${\bf I}$ is the identity matrix, $T_{ij}$ is the dipole-dipole
interaction tensor, and
$\omega_i$ and $\alpha_i$ are the characteristic frequency and
the static dipole polarizability, respectively, of the $i$-th oscillator.
The interaction (correlation) energy is given by the
difference between the ground state energy of the {\it coupled} system of
QHOs (proportional to the square root of the eigenvalues
{$\lambda_p$} of the $C^{QHO}$ matrix) and the ground state energy of the
{\it uncoupled} system of QHOs (derived from the characteristic frequencies):

\begin{equation}
E_{c,QHO} = 1/2 \sum_{p=1}^{3N} \sqrt{\lambda_p} - 3/2 \sum_{i=1}^{N}\omega_i
\,.
\label{interac}
\end{equation}

The so-computed interaction energy naturally includes many body
energy contributions, due to the dipole--dipole coupling among the QHOs;
moreover, it can be proved\cite{QHO} that 
the QHO model provides an efficient description of the
correlation energy for a set of localized fluctuating dipoles at an effective
Random Phase Approximation (RPA)-level. This is important because, 
differently from other schemes, RPA includes the effects of long-range
screening of the vdW interactions\cite{Goltl}, which are clearly 
of relevance, particularly for extended systems\cite{Klimes,Bucko,JCTC}.
The QHO interaction energy accounts for the long-range component of the
correlation energy, and is added to the energy computed within
the underlying semi-local DFT approximation.
Due to the short-range character of semi-local
functionals (PBE in our case), this procedure avoids double counting of
the correlation energy since a proper short-range damped form of the
QHO interaction is used.

The QHO model can be combined with the MLWF technique by 
assuming that each MLWF is represented by a three-dimensional isotropic
harmonic oscillator, so that
the system is described as an assembly of fluctuating dipoles.
By considering\cite{QHO} the Coulomb interaction 
between two spherical Gaussian charge densities
to account for orbital overlap at short distances (thus introducing
a short-range damping):

\begin{equation}
 V_{ij}=\frac{{\it erf}(r_{ij}/\sigma_{ij})} {r_{ij}}\;,
\label{Vij}
\end{equation}

where $r_{ij}$ is the distance between the $i$-th and the $j$-th 
Wannier Function Center (WFC), and  $\sigma_{ij}$ is an effective width,
$\sigma_{ij}=\sqrt{S_i^2 + S_j^2}$, where $S_i$ is the spread 
of the $i$-th MLWF, $w_i$; $S_i^2$ is defined as 
$\left<w_i|r^2|w_i\right> -\left<w_i|{\bf r}|w_i\right>^2 $.
Then, in Eq. \eqref{CQHO} the dipole interaction tensor is\cite{QHO}

\begin{equation}
T_{ij}^{ab} = -\frac{3r_{ij}^ar_{ij}^b-r_{ij}^2\delta_{ab}} {r_{ij}^5}
\left( {\it erf}\left(\frac {r_{ij}} {\sigma_{ij}}\right)-
\frac{2} {\sqrt{\pi}}\frac{r_{ij}} {\sigma_{ij}}
e^{-\left(\frac{r_{ij}} {\sigma_{ij}}\right)^2} \right) +
\frac{4} {\sqrt{\pi}}\frac{1} {\sigma_{ij}^3}
\frac{r_{ij}^ar_{ij}^b} {r_{ij}^2}
e^{-\left(\frac{r_{ij}} {\sigma_{ij}}\right)^2} 
\label{T}
\end{equation}

where $a$ and $b$ specify Cartesian coordinates ($x,y,z$), 
$r_{ij}^a$ and $r_{ij}^b$ are the respective components of the
distance $r_{ij}$, and $\delta_{ab}$ is the
Kronecker delta function.

Moreover, similarly to Eq. \eqref{alfa}, the polarizability is written as
                                                                                
\begin{equation}
\alpha_i = \zeta \frac{Z_i e^2}{m\omega_i^2}\,,
\label{alfabis}
\end{equation}

where, if spin degeneracy is exploited, $Z_i = 2$ since every MLWF corresponds
to 2 paired electrons.
Then, given again the direct relation between polarizability and 
volume, we assume that $\alpha_i = \gamma S_i^3$.

Similarly to refs. \cite{QHO-WF,QHO-WFs}, we combine 
the QHO model, which accurately describes the long-range
correlation energy, with a given semilocal, GGA functional (PBE in
our case), which is expected to well reproduce short-range correlation
effects, by introducing an empirical parameter $\beta$ that
multiplies the QHO-QHO parameter $\sigma_{ij}$ in Eq. \eqref{Vij}.
The three parameters $\beta$, $\gamma$, and $\zeta$ are set up by minimizing the
mean absolute relative errors (MARE), measured with respect
to high-level, quantum-chemistry reference values relative to the
S22 database of intermolecular interactions\cite{Jurecka}, a widely
used benchmark database, consisting of weakly interacting molecules
(a set of 22 weakly interacting dimers mostly of biological importance), 
with reference binding energies calculated
by a number of different groups using high-level
quantum chemical methods. 
By taking PBE as the reference DFT functional, we get: 
$\beta=1.39$, $\gamma=0.88$, and $\zeta=1.30$\cite{QHO-WF}.    
Once the $\gamma$ and $\zeta$ parameters are set up, 
both the polarizability $\alpha_i$ and the characteristic frequency 
$\omega_i$ are obtained 
just in terms of the MLWF spreads (see Eq. \eqref{alfabis} and below).

As in ref. \cite{QHO-WF}, in order to describe screening effects in the 
metal substrate, the potential of Eq. \eqref{Vij} is replaced by 
 
\begin{equation}
 V_{ij}=\frac{{\it erf}(r_{ij}/\sigma_{ij})\, e^{-q r_{ij}} } {r_{ij}}\;,
\label{VijTF}
\end{equation}

where $q$ is the standard Thomas-Fermi wave vector, $k_{TF}$, appropriate for
the substrate bulk metal if both
the $i$-th and the $j$-th WFC are inside the metal slab, $q=0$
if both the WFCs are outside the metal slab, while, in the
intermediate cases, $q=k_{TF}\, r_{ij}^{in}/r_{ij}$, that is
$k_{TF}$ is renormalized by considering the portion, $r_{ij}^{in}$, of the
$r_{ij}$ segment which is inside the metal slab.

In this way the method includes both a
short-range damping (to take orbital overlap effects into account) and
a long-range damping (to take metal-screening effects
into account).

\subsection{Computational details}
We here apply the DFT/vdW-WF2s1, DFT/vdW-WF2s3, and DFT/vdW-QHO-WF methods 
to the case of adsorption of Xe on the Ag(111), Au(111),
and Cu(111) metal surfaces.
All calculations have been performed 
with the Quantum-ESPRESSO ab initio package\cite{ESPRESSO} and the
MLWFs have been generated as a post-processing calculation using
the WanT package\cite{WanT}. Similarly to our previous 
studies\cite{Ar-Pb,PRB2013} we modeled the metal surface 
using a periodically-repeated hexagonal supercell, 
with a $(\sqrt{3}\times \sqrt{3})R30^{\circ}$ structure and a surface slab
made of 15 metal atoms distributed over 5 layers considering the 
experimental Ag(111), Au(111), and Cu(111) lattice constants. 
Repeated slabs were
separated along the direction orthogonal to the surface by a vacuum region
of about 20 \AA\ to avoid significant spurious interactions due to periodic
replicas. The Brillouin Zone has been sampled using a
$6\times6\times1$ $k$-point mesh. 
In this model system 
the coverage is 1/3, i.e. one adsorbed adatom for each 3
metal atoms in the topmost surface layer. 
The $(\sqrt{3}\times \sqrt{3})R30^{\circ}$
structure has been indeed observed\cite{Seyller} at low temperature by LEED for
the case of Xe adsorption on Cu(111) and Pd(111) (actually, 
this is the simplest commensurate structure for rare gas monolayers on 
close-packed metal surfaces and the only one for which good 
experimental data exist), and it was adopted in most of 
the previous ab initio 
studies\cite{Silva,DaSilva05,DaSilva,Righi,Lazic,Zhang}.
The metal surface atoms
were kept frozen (after a preliminary relaxation of the outermost
layers of the clean metal surfaces) and only the 
vertical coordinate (perpendicular to the surface) of the adatoms
was optimized, this procedure being justified
by the fact that only minor surface atom displacements are observed
upon physisorption\cite{DaSilva05,Zhang,Abad,Fajin}. 
Moreover, the adatoms
were adsorbed on both sides of the slab: in this way the surface dipole
generated by adsorption on the upper surface of the slab is cancelled by the
dipole appearing on the lower surface, thus greatly reducing the
spurious dipole-dipole interactions between the periodically repeated images 
(previous DFT-based calculations have shown
that these choices are appropriate\cite{DaSilva,Sun,Ar-Pb,Chwee}).

We have carried out calculations for various separations 
of the Xe atoms adsorbed on the {\it top} high-symmetry site
(on the top of a metal atom), since this is certainly the favored adsorption
site for Xe\cite{Ar-Pb,Lee2012,Liao,Michaelides,Hodgson}. 
For the Xe-Ag(111) system we have also considered adsorption on the 
{\it hollow} site (on the center of the triangle formed by the 3 surface 
metal atoms contained in the supercell) in order to verify whether the
present schemes are able to correctly predict which configuration
is energetically favored (see discussion in ref. \cite{Ar-Pb}).

In principle the adsorption on metal surfaces is challenging for a
Wannier-based scheme since in metal slabs the electronic charge is
relatively delocalized\cite{Heinrich} and the assumption of
exponential localization of the MLWFs is no longer strictly valid\cite{Resta}.
However, even in this case our methods perform well, as confirmed by the
fact that the spreads of our computed MLWFs are not larger than 2.5 \AA\
for the systems we have considered, so that the MLWFs are relatively
localized although the total electronic density is certainly not.
This does not come to a surprise;
in fact, on the one hand, the MLWF technique has been efficiently
generalized also to metals\cite{Souza,Iannuzzi}, on the other,
bonding in metallic clusters and in fcc bulk metals (like Ag, Au, and Cu)
can be described in terms of Hydrogen-like orbitals localized on tetrahedral
interstitial sites\cite{Souza}, which is just in line
with the spirit of our vdW-corrected schemes based on MLWFs.

For a better accuracy, as done in previous applications on adsorption
processes\cite{silvsurf,silvmetodo,silvinter,ambrosetti,Ar-Pb}, we have also
included the interactions of the MLWFs of the physisorbed fragments not 
only with the MLWFs of the underlying surface, within the reference supercell,
but also with a sufficient
number of periodically-repeated surface MLWFs (in any case, given the
$R^{-6}$ decay of the vdW interactions, the convergence with
the number of repeated images is rapidly achieved).
Electron-ion interactions were described using ultrasoft
pseudopotentials by explicitly including
11 valence electrons per Ag, Au, and Cu atom.
We chose the PBE\cite{PBE} reference DFT functional, which is probably
the most popular GGA functional.
The problem of choosing the optimal DFT functional, particularly in its
exchange component, to be combined
with long-range vdW interactions and the related problem of
completely eliminating double counting of correlation effects 
still remain open\cite{Riley}; however they
are expected to be more crucial for adsorption systems characterized
by relatively strong adparticle-substrate bonds (''chemisorption'') and,
for instance, for the determination of the perpendicular vibration
frequency\cite{Lazic} than for the equilibrium properties of the
physisorbed systems we focus on in our paper.

The additional cost of the post-processing vdW correction is basically
represented by the cost of generating the Maximally-Localized
Wannier functions from
the Kohn-Sham orbitals, which scales linearly with the size of the
system\cite{Marzari}. 

\section{Results and Discussion}
In Tables 1-3 and Fig. 1 we report results evaluated 
including the vdW corrections 
using our screened DFT/vdW-WF2s1 and
DFT/vdW-QHO-WF methods, and DFT/vdW-WF2, namely the unscreened version of
DFT/vdW-WF2s1. We also add data obtained by the 
simple, single-layer DFT/vdW-WF2s3 scheme (see Method section).
Our estimated binding energies and equilibrium distances are compared
(see Table 1) to the best available, reference experimental
and theoretical values, and to those obtained by other DFT
vdW-corrected schemes, including dispersion corrected PBE (PBE-D\cite{PBE-D}),
vdW-DF\cite{Dion,Langreth07}, vdW-DF2\cite{Lee-bis}, rVV10\cite{Sabatini},
and by the simpler Local Density Approximation (LDA) and semilocal GGA
(in the PBE flavor\cite{PBE}) approaches.
For the Xe-Ag(111) case we also report recent data computed by 
the PBE+vdW$^{surf}$\cite{Maurer}, PBE+MBD\cite{Maurer}, 
and cRPA+EXX\cite{Rohlfing} methods.

The {\it binding energy}, $E_b$, is defined as
\begin{equation}
E_b=1/2(E_{tot}-(E_s+2E_a))
\end{equation}
where $E_{s,a}$ represent the energies of the 
isolated fragments (the substrate and the adatoms)
and $E_{tot}$ is the energy of the interacting system, including the 
vdW-correction term (the factors 2 and 1/2 are due to the adsorption 
on both sides of the slab);
$E_s$ and $E_a$ are evaluated using the same supercell adopted
for $E_{tot}$.

$E_b$ has been evaluated for several adsorbate-substrate distances; then
the equilibrium distances and the corresponding binding energies 
have been obtained (as in refs. \cite{Ar-Pb,PRB2013}) by fitting
the calculated points with
the function: $A\,e^{-Bz}-C_3/(z-z_0)^3$
(plotted in Fig. 1 for the Xe-Ag(111) case; the Xe-Au(111) and 
the Xe-Cu(111) binding energies look very similar).
Typical uncertainties in the fit are of the order of $0.05$ \AA$ $ 
for the distances and a few meVs for the minimum binding energies.
For the adsorption of rare gases on the (111) noble-metal surface, 
reference data are available, particularly the "best estimates" reported by 
Vidali {\it et al.} \cite{Vidali}, that represent averages over different
theoretical and experimental evaluations.

As found in the previous studies\cite{Ar-Pb,PRB2013} 
the effect of the vdW-corrected schemes (see Table 1 and Fig. 1) is 
a much stronger bonding than with a pure PBE scheme (PBE yielding no 
significant binding), with the formation of a clear minimum in the
binding energy curve at a shorter equilibrium distance.
Moreover, by comparing with unscreened DFT/vdW-WF2 data 
we see that the effect of screening
is substantial, leading to reduced binding energies and 
increased adatom-substrate equilibrium distances:
the unscreened approach evidently overbinds.
With respect to the reference values, our screened methods appear to 
well reproduce the equilibrium binding energies,
although the equilibrium distances are slightly shorter.

All the considered theoretical schemes (see Table 2)
predict that the {\it top} site is favored with respect to the {\it hollow} 
one for Xe on Ag(111) (in agreement with the experimental 
evidence\cite{Diehl}), with the exception of PBE-D, thus confirming 
that employing a semiempirical approach is inappropriate for such a system.
PBE predicts that the {\it top} and {\it hollow} configurations are 
essentially isoenergetic since the equilibrium Xe-Ag(111) distance is
largely overestimated and the bonding strength largely underestimated.
Note also that with the unscreened DFT/vdW-WF2 approach the difference 
between the binding energies of the {\it top} and {\it hollow} structures
is substantially overestimated.
Interestingly, for the systems considered, the simple DFT/vdW-WF2s3 scheme,
based on the single-layer approximation also performs well.

The slight underestimate of our binding energies compared to the 
"best estimates" values\cite{Vidali} could also be rationalized by 
considering that the experimentally measured adsorption energy often
includes not only the interaction of adatoms with the substrate but
also lateral interactions among adatoms\cite{Sun,Ar-Pb}, thus leading
to higher energy estimates.

Concerning the other vdW-corrected methods considered,
PBE-D gives reasonable equilibrium distances but largely overbinds
(much more than LDA), particularly for the Xe-Au(111) where the error 
is of more than 100 \%. This bad performance is not unexpected since 
such a semiempirical, atom-based approach cannot well describe metal 
substrates with their delocalized electronic charge and screening effects. 
rVV10 gives reasonable equilibrium distances but tends to overbind
(more than LDA), while vdW-DF and vdW-DF2,
as expected\cite{Klimes} for this kind of vdW functionals, 
substantially overestimate the equilibrium 
adsorption distances, with vdW-DF2 which also turns out to significantly
underestimate the binding energy.  

From Table 1 one can also see that the binding energies
are reasonably reproduced by the LDA scheme for the all cases,
a behavior common to several physisorption systems.
However, as already outlined above,
this agreement should be considered accidental: the well-known
LDA overbinding, due to the overestimate of the long-range part of the
exchange contribution, somehow mimics the missing vdW interactions;
moreover the equilibrium distances predicted by LDA are
clearly underestimated since LDA
cannot reproduce the $R^{-6}$ behavior in the interaction potential,
so that the binding energy exhibits a wrong asymptotic behavior at a large
distance $Z$ from the surface (decaying exponentially rather 
than as $\sim 1/Z^3$).

Looking at data for the Xe-Ag(111) system, one can also see the
relatively good performances of the PBE+vdW$^{surf}$ and PBE+MBD
methods\cite{Maurer}.
The underbinding exhibited by the cRPA-EXX calculation of 
ref. \cite{Rohlfing}, which combines exact exchange and random-phase 
approximation (RPA) correlation without using any density functional, 
is probably due\cite{Maurer} to 
neglect of the exchange-correlation kernel, the underlying 
plasmon-pole approximation, and the fact that the response function of 
the system is not fully coupled, being calculated separately for substrate 
and adsorbate.

As can be seen from Table 1, the values of the binding energies of 
Xe on Ag(111), Au(111),
and Cu(111) are comparable, with Xe-Au(111) which turns out to be the
system which is energetically slightly more favored according to most 
of the vdW-corrected schemes. Interestingly, using the unscreened
DFT/vdW-WF2 method the Xe-Au(111) is instead the least energetically 
system, thus confirming once again the importance of screening
effects in adsorption processes on metal surfaces.

Concerning the computed $C_3$ coefficients (see Table 3) using
data obtained by our methods and fitting the binding-energy curve, they
appear in reasonable, semiquantitative agreement with reference 
data (particularly for the Xe-Ag(111) system), although one should 
remember that our estimated values cannot be very accurate given 
the limited size of our simulation slab and vacuum thickness
(as discussed, for instance, in ref. \cite{Rohlfing}).

\section{Conclusions}
In summary, we have investigated the adsorption of Xe
on the Ag(111), Au(111), and Cu(111) metal surfaces, 
by considering our screened DFT/vdW-WF2s1, 
DFT/vdW-WF2s3, and DFT/vdW-QHO-WF methods.
By analyzing the results of our study and comparing them to
available reference data, we get a substantial improvement
with respect to the unscreened DFT/vdW-WF2 approach.
Given the uncertainties in the reference data, one cannot
easily state which scheme is more appropriate.
Considering all the studied cases DFT/vdW-QHO-WF turns out
to be marginally superior which correlates with the relatively
higher complexity of this approach.
Interestingly, we confirm the conclusion of previous studies
(see, ref. \cite{Hanke} and references therein) which suggest that, 
particularly for the close-packed (111) surfaces, the assumption of a 
one-layer screening depth (single-layer approximation) works
reasonably well (DFT/vdW-WF2s3 approach).
The differences between the values of the equilibrium binding energies
and distances predicted by our adopted different schemes can be
taken as the order of magnitude of the uncertainty associated to
the different screened methods and to estimate their accuracy. 

For the considered systems, in general our methods are comparable 
with the most recent vdW-corrected schemes, such as PBE+vdW$^{surf}$ and
PBE+MBD; moreover they perform better than the
semiempirical PBE-D method and the
popular vdW-DF and vdW-DF2 approaches, which, in particular,
exhibit a general tendency to overestimate the equilibrium distances,
in line with the behavior reported for systems
including a metallic surface\cite{Vanin}.


\vfill
\eject

\begin{table}
\caption{Binding energy per Xe atom E$_b$ (in meV) and (in parenthesis) 
equilibrium distance R (in \AA) for Xe adsorbed on Ag(111), Au(111), and
Cu(111) metal surfaces in the {\it{top}} configuration, 
using different methods. Results are compared with available reference data.}
\begin{center}
\begin{tabular}{|l|c|c|c|}
\hline
method            &  Xe-Ag(111)  &  Xe-Au(111)  &  Xe-Cu(111)   \\ \tableline
\hline
LDA               & -215 (3.12)   & -230 (3.14) & -204 (3.04) \\
PBE               &  -18 (5.02)   &  -19 (4.34) &  -19 (4.49) \\
\hline
PBE-D             & -282 (3.54)   & -517 (3.22) & -272 (3.40) \\
rVV10             & -236 (3.53)   & -280 (3.48) & -223 (3.47) \\
vdW-DF            & -180 (4.08)   & -199 (4.00) & -184 (3.97) \\
vdW-DF2           & -154 (4.00)   & -180 (3.86) & -157 (4.01) \\
\hline
PBE+vdW$^{surf}$  & -220 (3.56)$^a$&      ---    &     ---     \\ 
PBE+MBD           & -170 (3.64)$^a$&      ---    &     ---     \\ 
cRPA+EXX          & -140 (3.60)$^b$&      ---    &     ---     \\
\hline
DFT/vdW-WF2       & -298 (3.13)   & -277 (3.36) & -304 (3.11) \\
DFT/vdW-WF2s1     & -186 (3.34)   & -209 (3.42) & -189 (3.24) \\
DFT/vdW-WF2s3     & -179 (3.37)   & -197 (3.44) & -197 (3.23) \\
DFT/vdW-QHO-WF    & -199 (3.45)   & -227 (3.49) & -188 (3.28) \\ 
\hline
reference  &-230$\leftrightarrow$-180 (3.45$\leftrightarrow$3.68)$^c$&-214$^d$&-280$\leftrightarrow$-183 (3.20$\leftrightarrow$4.00)$^{d,e}$ \\
"best estimate"$^d$&-211$\pm 15$(3.60$\pm 5$) &  --- & -183$\pm 10$ (3.60) \\
\hline
\end{tabular}
\tablenotetext[1]{ref.\cite{Maurer}.}
\tablenotetext[2]{ref.\cite{Rohlfing}.}
\tablenotetext[3]{ref.\cite{refAg}.}
\tablenotetext[4]{ref.\cite{Vidali}.}
\tablenotetext[5]{refs.\cite{Seyller,Lazic,Chen}.}
\end{center}
\label{table1}
\end{table}
\vfill
\eject

\begin{table}
\caption{Binding energy per Xe atom E$_b$ (in meV) and (in parenthesis) 
equilibrium distance R (in \AA) for Xe adsorbed on Ag(111) in the
{\it{top}} and {\it{hollow}} configurations, using different methods.}
\begin{center}
\begin{tabular}{|l|c|c|}
\hline
method            &   {\it{top}}  & {\it{hollow}}\\ \tableline
\hline
LDA               & -215 (3.12)   & -205 (3.07)  \\
PBE               &  -18 (5.02)   &  -18 (4.22)  \\
\hline
PBE-D             & -282 (3.54)   & -317 (3.38)  \\
rVV10             & -236 (3.53)   & -234 (3.52)  \\
vdW-DF            & -180 (4.08)   & -179 (4.08)  \\
vdW-DF2           & -154 (4.00)   & -152 (4.01)  \\
\hline
DFT/vdW-WF2       & -298 (3.13)   & -246 (3.24)  \\
DFT/vdW-WF2s1     & -186 (3.34)   & -175 (3.31)  \\
DFT/vdW-WF2s3     & -179 (3.37)   & -167 (3.40)  \\
DFT/vdW-QHO-WF    & -199 (3.45)   & -198 (3.46)  \\ 
\hline
\end{tabular}
\end{center}
\label{table2}
\end{table}
\vfill
\eject

\begin{table}
\caption{Estimated $C_3$ coefficients, in eV\AA$^3$, for Xe
in the {\it top} configuration on the metal surfaces computed using 
data obtained by our methods (by fitting the binding-energy curve, 
see text), compared to available reference data.}
\begin{center}
\begin{tabular}{|l|c|c|c|}
\hline
method              & Xe-Ag(111)&Xe-Au(111)&Xe-Cu(111)  \\ \tableline
\hline
DFT/vdW-WF2         & 3.29 & 5.59 & 5.84 \\
DFT/vdW-WF2s1       & 3.30 & 4.30 & 4.25 \\
DFT/vdW-WF2s3       & 3.41 & 4.11 & 4.38 \\
DFT/vdW-QHO-WF      & 4.62 & 5.59 & 4.92 \\ 
\hline
reference           $^a$& 3.28 &3.20$\leftrightarrow$3.79& 3.39 \\  
\hline
\end{tabular}
\tablenotetext[1]{ref.\cite{Vidali}.}       
\end{center}
\label{table3}
\end{table}
\vfill
\eject

\pagestyle{empty}
                      
\begin{figure}
\centerline{
\includegraphics[width=17cm]{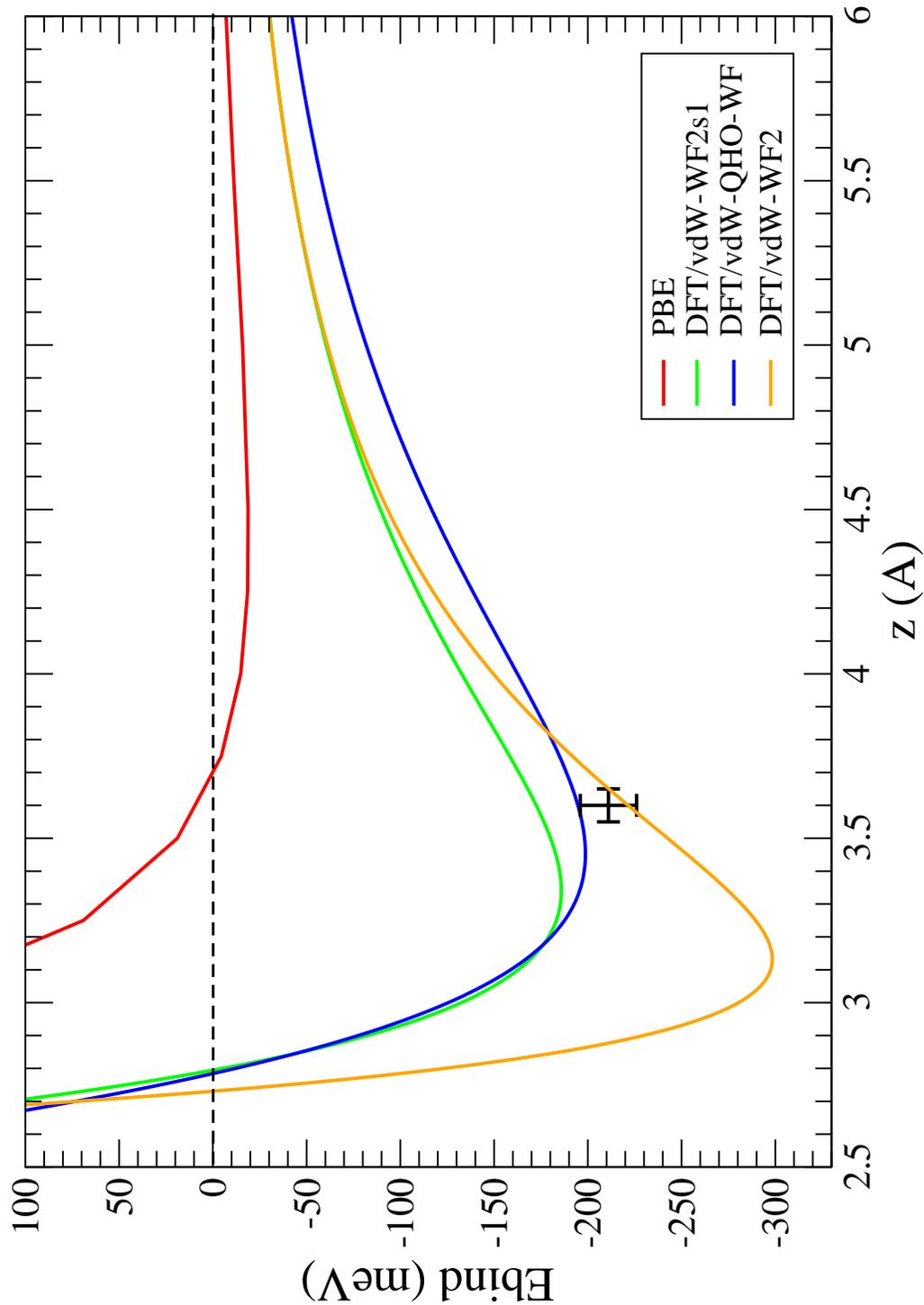}
}
\caption{Binding energy of Xe on Ag(111), 
as a function of the distance from the Ag(111) surface,
computed using the standard PBE
calculation, and including the vdW corrections using our 
(unscreened) DFT/vdW-WF2, and (screened) DFT/vdW-WF2s1, and DFT/vdW-QHO-WF
methods; the position of the "best estimate" value\cite{Vidali} with error
bars is also reported.} 
\label{fig1}
\huge
\end{figure}
                      
\end{document}